\documentclass{ws-ijmpa}
\usepackage[compress]{cite}
\usepackage{graphicx}

\begin{document}
\title{Quantum Entanglement and Thermodynamics of Boson-Antiboson Pair Creation Modes in the Noncommutative Bianchi I Universe}
\author{M. F. Ghiti$^{1,2}$, N. Mebarki$^{2}$ and H. Aissaoui$^{2}$}
\address{1. Ecole Normale Sup\'{e}rieure (ENSC), Assia Djebar, Constantine, Algeria\\
2. Laboratoire de Physique Math\'{e}matique et Subatomique (LPMS),\\ 
Constantine 1 University, Constantine, Algeria \\farouk.ghiti@yahoo.com}

\maketitle
\begin{abstract}
Within the quantum field theory approach and using the technique of Bogoliubov transformations, 
the von Neumann boson-antiboson pair creation quantum 
entanglement entropy is studied in the context of the noncommutative Bianchi I universe. It is shown that the latter have a 
behavior depending strongly on the choice of the noncommutativity 
$\theta$ parameter, $k_{\perp}$-frequency modes and the structure of the curved spacetime. 
Moreover, the relationship between the Bose-Einstein condensation and quantum entanglement is also discussed.    
\end{abstract}

\ccode{PACS numbers: 03.65.Ud, , 03.65.-w, 03.67.-a, 03.75.Nt, 04.62.+v}

%%%%%%%%%%%%%%%%%%%%%%%%%%%%%%%%%%%%%%%%%%%%%%%%%%%%%%%%%%%%%%%%%%%%%%%%%%%%%%%%%%%%%%%%%%%%%%%%%%%%%%%%%%%%%%%%%%%%%%%%%%%%%%%%
\section{Introduction}
\label{Int}
During the last few years, The quantum entanglement (Q.E.) phenomenon has been 
of great interest for many people in the field of quantum information in both flat and 
curved spacetime \cite{1,2,3,4,5,6,7,8,9,10,11,12,13,14}. Moreover, it turns out that 
the pair creation process is intimately related to the quantum entanglement and strongly depends 
on the nature of the particles as well as on the structure of the spacetime \cite{5,9,11,15,16}. 
Furthermore, as a possible scenario to explain dark matter and dark energy as well as the anisotropies 
found in the cosmic microwave background (C.M.B.), is the noncommutative geometry (N.C.G) \cite{17,18,19,20,21,22}. 
In fact, many models have been proposed so far where it has been shown that the noncommutativity 
$\theta$-parameter plays an important role \cite{23,24,25,26,27,28,29,30,31,32}. 
The goal of this paper is to study the process of creating boson-antiboson pairs in the noncommutative (N.C.)
Bianchi I universe and show its relationship with quantum entanglement and Bose-Einstein condensation
(B.E.C.) phenomena. The role played by the spacetime noncommutativity is also emphasized. 
In section 2, we present the necessary mathematical formalism. Section 3 contains our numerical 
results and discussions. Finally, in section 4, we draw our conclusions. 
%%%%%%%%%%%%%%%%%%%%%%%%%%%%%%%%%%%%%%%%%%%%%%%%%%%%%%%%%%%%%%%%%%%%%%%%%%%%%%%%%%%%%%%%%%%%%%%%%%%%%%%%%%%%%%%%%%%%%%%%%%%%%%%%
\section{Mathematical Formalism}
\label{math}
The N.C. spacetime is characterized by the coordinates operators $
\hat{x}^{\mu}$ $(\mu=\overline{0,3})$ satisfying the following commutation 
relation:
\begin{equation}
\left[ \hat x^{\mu},\hat x^{\nu}\right]=i \,\theta^{\mu\nu}
\end{equation}
where $\theta^{\mu\nu}$ are antisymmetric matrix elements which control the 
noncommutativity of the spacetime. The N.C. matter scalar density of the massless Seiberg-Witten (S.W.) bosonic field $\hat{\varphi}$ reads :
\begin{equation}
\mathcal{L} = \hat{e} \ast \left(\hat{g}^{\mu\nu}\ast\left(\hat{D}_{\mu}\hat{\varphi}\right)^{\dagger}\ast\hat{D}_{\nu}\hat{\varphi}\right)
\end{equation}
here, the deformed tetrad is given by : 
\begin{equation}
\hat{e} =det_{\ast}\left(\hat{e}^{a}_{\mu}\right)=\frac{1}{4!}\epsilon^{\mu\nu\rho\sigma}\epsilon_{a\,b\,c\,d}\,\hat{e}^{a}_{\mu}\ast\hat{e}^{b}_{\nu}\ast\hat{e}^{c}_{\rho}\ast\hat{e}^{d}_{\sigma}
\end{equation}
where $\epsilon^{\mu\nu\rho\sigma}$ and $\epsilon_{a\,b\,c\,d}$ are the completely antisymmetric tensors in curved and flat spacetime respectively. The gauge covariant derivative is given by : $\hat{D}_{\mu}\hat{\varphi}=\left(\partial_{\mu}-ie\hat{A}_{\mu}\right)\ast\hat\varphi$. The corresponding Klein-Gordon equation is shown to take the following form (See Appendix A) : 
%(see \ref{appendix A}) \cite{28}:
\begin{eqnarray}\label{NC Klein}
\left(g^{\mu\nu}\partial_{\mu}\partial_{\nu}-\frac{1}{\sqrt{-g}}\partial_{\mu}\left(\sqrt{-g}g^{\mu\nu}\right)
\partial_{\nu}\right)\hat{\varphi}-\frac{i}{2\sqrt{-g}}\theta^{\alpha\beta}\partial_{\mu}\left(\partial_{\alpha}+
\sqrt{-g}\partial_{\beta}g^{\mu\nu}\partial_{\nu}\hat{\varphi}\right)+\notag\\\frac{1}{8\sqrt{-g}}\theta^{\alpha\beta}\partial_{\mu}\left(\partial_{\alpha}\partial_{\rho}\left(\sqrt{-g}g^{\mu\nu}\right)\partial_{\beta}\partial_{\sigma}\partial_{\nu}\hat{\varphi}+\partial_{\rho}\left(\partial_{\alpha}\sqrt{-g}\partial_{\beta}g^{\mu\nu}\partial_{\sigma}\partial_{\nu}\hat{\varphi}\right)+\right.\notag\\\left.\partial_{\alpha}\partial_{\rho}\sqrt{-g}\partial_{\beta}\partial_{\sigma}g^{\mu\nu}\partial_{\nu}\hat{\varphi}\right)+\frac{i}{2\sqrt{-g}}\theta^{\alpha\beta}\partial_{\mu}\left(\partial_{\alpha}\left(\sqrt{-g}g^{\mu\nu}\right)\partial_{\beta}\partial_{\nu}\hat{\varphi}\right)=0
\end{eqnarray}
where $g$ is the determinant of the metric.
In what follows, using dimensionless spacetime coordinates $(t, x, y , z)$, we consider a Bianchi I universe where the metric has the form:
\begin{equation}\label{metric}
ds^{2}=-dt^{2}+t^{2}\left( dx^{2}+dy^{2}\right)+dz^{2}
\end{equation}
(here the time $t$ is related to the cosmological parameters of the model). By convention, we choose the spacetime signature as being $(-,+,+,+)$ and for simplicity, we take :
\begin{equation}
\theta_{\mu\nu}=\left[
\begin{array}{cccc}
0 & 0 & \theta & 0\\
0 & 0 & 0 & 0\\
-\theta & 0 & 0 & 0\\
0 & 0 & 0 & 0
\end{array}\right]
\end{equation}
(notice that the choice of the $\theta^{\mu\nu}$ does not affect so much our conclusions). Using Maple 16 tensor package, the non-vanishing components of the vierbeins up to $O\left(\theta^{2}\right)$ have the following expressions:
\begin{eqnarray}\label{vierbeins}
\hat e^{\tilde{0}}_{0}&=&\hat e^{\tilde{3}}_{3}=1\notag\\
\hat e^{\tilde{1}}_{1}&=&\hat e^{\tilde{2}}_{2}=\frac{1}{t}\Big(1-
\frac{25\,\theta^{2}}{128}\Big)
\end{eqnarray}
and therefore, the corresponding deformed Bianchi I metric up to $O\left(\theta^{2}\right)$ is given by:
\begin{equation}\label{NC metric}
ds^{2}=-dt^{2}+t^{2}\left(1-\frac{21}{64}\theta^{2}\right)\left(dx^{2}+dy^{2}\right)+dz^{2}
\end{equation}
It is worth mentioning, as pointed out in references \cite{5,29,33,34,35} that since the metric 
presents a space-like singularity at $t = 0$, it is difficult to define the particle state within the
 adiabatic approach. To do so, we first follow a quasiclassical approach of Ref. \citen{34} to identify the positive 
and negative frequency modes and look for the asymptotic behavior of 
the solutions at $t\rightarrow 0$ and $t\rightarrow\infty$. Secondly, we solve the N.C. Klein Gordon 
equation and compare the solutions with the above quasiclassical limit. After straightforward but lengthy calculations, Eq. (\ref{NC Klein}) is simplified to 
\begin{eqnarray}\label{Klein_calc}
\left(1+\frac{i\theta}{t}\tilde{\partial}_{2}-\frac{\theta^{2}}{4t^{2}}\tilde{\partial}_{2}^{2}\right)\tilde{\partial}_{0}^{2}\hat{\varphi}
+&&\frac{2}{t}\left(1+\frac{i\theta}{2t}\tilde{\partial}_{2}\right)\tilde{\partial}_{0}\hat{\varphi}
+\left(\frac{1}{t^{2}}\left(1+\frac{21}{64}\theta^{2}\right)\left(\tilde{\partial}_{1}^{2}+\tilde{\partial}_{2}^{2}\right)+\right.\notag\\
&&\left.\left(1+\frac{i\theta}{t}\tilde{\partial}_{2}-\frac{\theta^{2}}{4t^{2}}\tilde{\partial}_{2}^{2}\right)\tilde{\partial}_{3}^{2}\right)\hat{\varphi}=0
\end{eqnarray}
here, the tilde stands for a curved spacetime index. To simplify this equation, we set: 
\begin{equation}
\hat{\varphi}=\hat{t}^{-1}f\left(t\right)h\left(t\right)\exp[i\left(k_{x}x+k_{y}y+k_{z}z\right)]
\end{equation} 
where
\begin{equation}
\hat{t}=t-\frac{\theta}{2}k_{y}, \qquad f\left(t\right)=\exp\left(-\frac{\theta}{2t}k_{y}\right)
\end{equation}
where $h\left(t\right)$ is some regular function. After straightforward but tedious calculations Eq. (\ref{Klein_calc}) becomes: 
\begin{equation}\label{Klein_simplified}
\left(\frac{d^{2}}{dt^{2}}+\frac{{k^{2}_{\perp}}+\frac{21}{64}\theta^{2}{k^{2}_{\perp}}}{t^{2}}+k^{2}_{z}\right)h\left(t\right)=0
\end{equation}
where $k_{\perp}$ is given by:
\begin{equation}
k^{2}_{\perp}=k^{2}_{x}+k^{2}_{y}
\end{equation}
Now, we adopt the following change of variables :
\begin{subequations}
\begin{align}
&h\left(t\right)=\rho^{\mu_{\theta}+\frac{1}{2}}\exp\left({-\frac{\rho}{2}}\right)y\left(\rho\right)&\\
&\rho=-2ik_{z}t&\\ 
&\mu_{\theta}=i\sqrt{-\frac{1}{4}+k^{2}_{\perp}+\frac{21}{64}\theta^{2}k^{2}_{\perp}}&
\end{align}
\end{subequations}
Notice that after direct calculations, Eq.(\ref{Klein_simplified}) takes the following form:
\begin{equation}\label{Kummer}
\left(\rho\frac{d^{2}}{d\rho^{2}}+\left(2\mu_{\theta}+1-\rho\right)\frac{d}{d\rho}-\left(\mu_{\theta}+\frac{1}{2}\right)\right)y\left(\rho\right)=0
\end{equation}
It is important to mention that Eq. (\ref{Kummer}) has the form of the Kummer's differential equation \cite{5,36} : 
\begin{equation}
\left(\rho\frac{d^{2}}{d\rho^{2}}+\left(b-\rho\right)\frac{d}{d\rho}-a\right)y\left(\rho\right)=0
\end{equation}  
with : 
\begin{subequations}
\begin{align}
&a=\mu_{\theta}+\frac{1}{2}&\\
&b=2\mu_{\theta}+1&
\end{align}
\end{subequations}  
The solution of Eq.(\ref{Kummer}) can be expressed as a linear combination of the 
two independent Kummer functions $M\left(\mu_{\theta}+\frac{1}{2},2\mu_{\theta}+1,\rho\right)$ and $U\left(\mu_{\theta}+\frac{1}{2},2\mu_{\theta}+1,\rho\right)$. 
Thus, the general solution of Eq. (\ref{Klein_simplified}) is given by : 
\begin{equation}\label{general_solution}
h\left(\rho\right)=\rho^{\mu_{\theta}+\frac{1}{2}}\exp\left({-\frac{\rho}{2}}\right)\left(C_{1}M\left(\mu_{\theta}+\frac{1}{2},2\mu_{\theta}+1,\rho\right)+C_{2}U\left(\mu_{\theta}+\frac{1}{2},2\mu_{\theta}+1,\rho\right)\right)
\end{equation}
($C_{1}$ and $C_{2}$ are normalization constants). Now, for a better understanding 
of the asymptotic behavior at $\rho\to 0$ ($"in"$ fields) and $\rho\to \infty$ ($"out"$ fields) of the solutions, 
Eq. (\ref{general_solution}), it is preferable to express the $M\left(\mu_{\theta}+\frac{1}{2},2\mu_{\theta}+1,\rho\right)$ 
and $U\left(\mu_{\theta}+\frac{1}{2},2\mu_{\theta}+1,\rho\right)$ Kummer functions in terms of Whittaker ones, i.e. \cite{36} :
 \begin{subequations}
\begin{align}
&M\left(\frac{1}{2}+\mu-\lambda,1+2\mu,z\right) = e^{\frac{z}{2}}\,z^{-\left(\frac{1}
{2}+\mu\right)}\,M_{\lambda,\mu}\left(z\right)&\\
&U\left(\frac{1}{2}+\mu-\lambda,1+2\mu,z\right) = e^{\frac{z}{2}}\,z^{-\left(\frac{1}
{2}+\mu\right)}\,W_{\lambda,\mu}\left(z\right)&
\end{align}
\end{subequations}  
Note that the Whittaker function $W_{\lambda,\mu}\left(z\right)$ can be expressed in terms of $M_{\lambda,\mu}\left(z\right)$ as follows :
\begin{equation}\label{Whittaker asymptotic}
W_{\lambda,\mu}\left(z\right)=\frac{\Gamma\left(-2\mu\right)}
{\Gamma\left(\frac{1}{2}-\mu-\lambda\right)}M_{\lambda,\mu}\left(z\right)+\frac{\Gamma\left(2\mu\right)}
{\Gamma\left(\frac{1}{2}+\mu-\lambda\right)}M_{\lambda,-\mu}\left(z\right)
\end{equation}  
where :
\begin{equation}
\left(W_{\lambda,\mu}\left(z\right)\right)^{\ast}=W_{-\lambda,\mu}\left(-z\right)
\end{equation}
and :
\begin{equation}
\left(M_{\lambda,\mu}\left(z\right)\right)^{\ast}=\left(-1\right)^{\mu+\frac{1}{2}}M_{\lambda,-\mu}\left(z\right)
\end{equation}
To construct the positive and negative frequency modes in the $"in"$ and $"out"$ fields, we have to use the asymptotic 
behavior of the Whittaker functions. It is easy to show that at $t\rightarrow 0 $ $\left(\rho\rightarrow 0\right)$ one has :
\begin{subequations}
\begin{align}
&h_{in}^{+}\sim M_{\lambda,\mu}\left(\rho\right)\sim e^{-\frac{\rho}{2}}\,\rho^{\mu+\frac{1}{2}}&
\\
&h_{in}^{-}\sim (M_{\lambda,\mu}\left(\rho\right))^{\star}\sim (-1)^{\mu+\frac{1}{2}}\,
M_{\lambda,-\mu}\left(\rho\right)&
\end{align}   
\end{subequations}    
Similarly, for $t\rightarrow\infty$ $\left(\rho\rightarrow \infty\right)$, the corresponding positive and negative frequency modes are :
\begin{subequations}
\begin{align}
&h_{out}^{+}\sim W_{\lambda,\mu}\left(\rho\right)\sim e^{-\frac{t}{2}}\,\rho^{\lambda}&\\
&h_{out}^{-}\sim (W_{\lambda,\mu}\left(\rho\right))^{\star}\sim W_{-\lambda,\mu}\left(-\rho\right)&
\end{align}   
\end{subequations}
Furthermore, the Bogoliubov coefficients associated with the asymptotic solutions $"in"$ and $"out"$ 
(involving in the past and the future respectively) read :
\begin{equation}\label{field theory}
h^{\pm}_{out}\left(t\right)=\alpha\,h^{\pm}_{in}\left(t\right)+\beta\,\left(h^{\pm}_{in}\left(t\right)\right)^{\ast}
\end{equation}
where the signs $"+"$ and $"-"$ stand for positive and negative frequency modes respectively. We  note that $\left(h^{\mp}_{in}\left(t\right)\right)^{\ast}=h^{\pm}_{in}\left(t\right)$. 
Comparing Eq. (\ref{Whittaker asymptotic}) with the Eq. (\ref{field theory}), we deduce that the Bogoliubov coefficients read :
\begin{equation}
\gamma=\frac{\beta}{\alpha}=\frac{-i\frac{\Gamma\left(2\mu_{\theta}\right)}{\Gamma\left(\frac{1}{2}+\mu_{\theta}-\lambda\right)}}{\frac{\Gamma\left(-2\mu_{\theta}\right)}{\Gamma\left(\frac{1}{2}-\mu_{\theta}-\lambda\right)}}\exp\left(-\pi\mu_{\theta}\right)
\end{equation} 
These coefficients for scalar particles satisfy the following normalization relation :
\begin{equation}
\vert\alpha\vert^{2}-\vert\beta\vert^{2}=1
\end{equation}
it is very important to rewrite the Eq. (\ref{field theory}) as :
\begin{equation}\label{phi equation}
\hat{\varphi}^{\pm}_{out}\left(t,x\right)=\alpha\,\hat{\varphi}^{\pm}_{in}\left(t,x\right)+\beta\left(\hat{\varphi}^{\pm}_{in}\left(t,x\right)\right)^{\ast}
\end{equation}
Using the fact that :
\begin{equation}\label{first relation}
\hat{\varphi}_{in,out}\left(t,\vec{r}\right)=\int\frac{d^{3}k}{\left(2\pi\right)^{3/2}}
\left[a_{in,out}\left(k\right)\exp\left(-i\,k\,X\right)+b^{+}_{in,out}\left(-k\right)\exp\left(i\,k\,X\right)\right]
\end{equation}
where $X=\left(\vec{r},t\right)$ and $k=\left(\vec{k},k^{0}\right)$ are respectively the four Minkowski position and momentum vectors. 
Here $a_{in,out}$ and $b^{+}_{in,out}$ are the 
annihilation and creation operators for boson and antiboson respectively.
By comparing Eq. (\ref{first relation}) with Eq. (\ref{phi equation}), we can deduce that :
\begin{equation}
b_{out}\left(-k\right)=\alpha\,a_{in}\left(k\right)+\beta\,b_{in}^{+}\left(-k\right)
\end{equation} 
Due to the form of the Bogoliubov transformations, we can show easily that the tensor product of the boson 
and antiboson vacuum state $\vert0\rangle_{out}\otimes\vert0\rangle_{out}=\vert 0,0\rangle_{out}$ can be 
written in terms of  the $"in"$ states as (Schmidt decomposition)\cite{16} :     
\begin{equation}
\vert0\rangle_{out}\otimes\vert0\rangle_{out}=\sum_{n=0}C_{n}\vert n_{k}\rangle_{in}\otimes\vert n_{-k}\rangle_{in}
\end{equation} 
where $\vert n_{k}\rangle_{in}$ and $\vert n_{-k}\rangle_{in}$ represent respectively the particle and the antiparticle mode 
states with momentum $k$ and $-k$. To find the relationship between the $C_{n}$ coefficients, 
we impose first the relation $b_{out}\left(-k\right)\vert 0,0\rangle_{out}=0$ leading to :
\begin{equation}
C_{n+1}=-\gamma\,C_{n-1}
\end{equation}
or :
\begin{equation}\label{cn}
\vert C_{n}\vert^{2}=\vert\gamma\vert^{2n}\vert C_{0}\vert^{2}
\end{equation}
Then, using the normalization condition for the vacuum state $\langle 0,0\vert 0,0\rangle_{out} = 1$ or equivalently  
\begin{equation}
\sum_{n=0}\vert C_{n}\vert^{2}=1
\end{equation}
and the fact that :
\begin{equation}
1+\vert\gamma\vert^{2}+\vert\gamma\vert^{4}+...=\frac{1}{1-\vert\gamma\vert^{2}}
\end{equation}
we end up with :   
\begin{equation}\label{c0}
\vert C_{0}\vert^{2}=1-\vert\gamma\vert^{2}
\end{equation}
Regarding the von Neumann quantum entanglement entropy $S_{Q.E}$ and since we are dealing with the bipartite pure state, then $S_{Q.E}$ can be written as : 
\begin{equation}
S_{Q.E}=-Tr \left(\rho\log_{2}\rho\right)=-\sum_{n=0}\lambda_{n}\log_{2}\lambda_{n}
\end{equation}
where $\rho$ is the density matrix, $"Tr"$ stands for trace and $\lambda_{n}$ are the eigenvalues of $\rho$ 
which are shown to be equal to $\vert C_{n}\vert^{2}$. Using Eq. (\ref{cn}) and Eq. (\ref{c0}) together with :
\begin{equation}
\sum n\,x^{n}=\frac{x}{\left(1-x\right)^2}
\end{equation}
 and $\sum_{n} x^{n}=\frac{1}{1-x}$, we obtain :
 \begin{equation}
 S_{Q.E}=\log_{2}\left(\frac{\vert\gamma\vert^{\left(2\vert\gamma\vert^{2}\right)/\left(\vert\gamma\vert^{2}-1\right)}}{1-\vert\gamma\vert^{2}}\right)
 \end{equation} 
 such that :
 \begin{equation}
 \vert\gamma\vert^{2}=\Big\vert\frac{\beta}{\alpha}\Big\vert^{2}=\Bigg\vert\frac{\Gamma\left(\frac{1}{2}-
\mu_{\theta}-\lambda\right)}{\Gamma\left(\frac{1}{2}+\mu_{\theta}-\lambda\right)}\Bigg\vert^{2}\exp\left(-2\pi\vert\mu_{\theta}\vert\right)
 \end{equation}
By using the relation :
 \begin{equation}
\Big\vert\Gamma\left(\frac{1}{2}+i\,y\right)\Big\vert^{2}=\frac{\pi}{\cosh\left(\pi\,y\right)}
 \end{equation}
 and after straightforward and tedious calculations, we get :
 \begin{equation}
 \Big\vert\frac{\beta}{\alpha}\Big\vert^{2}=\exp\left(-2\pi\vert\mu_{\theta}\vert\right)
 \end{equation}

%%%%%%%%%%%%%%%%%%%%%%%%%%%%%%%%%%%%%%%%%%%%%%%%%%%%%%%%%%%%%%%%%%%%%%%%%%%%%%%%%%%%%%%%%%%%%%%%%%%%%%%%%%%%%%%%%%%%%%%%%%%%%%%%
\section{Numerical Results and Discussions} 
\label{Res}
The numerical results of the Q.E. entropy $"S_{Q.E.}"$ in the N.C. Bianchi I universe show that it is a 
decreasing function of the $k_{\perp}-$frequency modes for a fixed value of the N.C. $\theta$ parameter. 
For example for $\theta=0.4$, when $k_{\perp}=1.0$, the $S_{Q.E.}=0.033$ and for $k_{\perp}=1.4$ for 
the same value of $\theta$, the $S_{Q.E.}=0.002$. Notice that contrary to the fermionic case of 
Ref. \citen{5}, $k_{\perp}-$ frequency modes does not start from the value $k_{\perp}=0$, for 
example, for $\theta =0$, $k_{\perp}^{min}=0.5$. As a function of $\theta$, one can show easily 
that $k_{\perp}^{min}=\frac{4}{\sqrt{25\theta^{2}+64}}$. The reason lies to the existence of the 
factor $\mu_{\theta}=\frac{1}{8}\sqrt{\left(25\theta^{2}+64\right)k_{\perp}^{2}-16}$ in the expression of the $S_{Q.E.}$ such that :
\begin{equation}
S_{Q.E}=\log_{2}\left(\frac{\exp\left(-2\pi\vert\mu_{\theta}\vert\right)^{\exp\left(-2\pi\vert\mu_{\theta}\vert\right)/
\left(\exp\left(-2\pi\vert\mu_{\theta}\vert\right)-1\right)}}{1-\exp\left(-2\pi\vert\mu_{\theta}\vert\right)}\right)
\end{equation} 
which has to be real. \\
\begin{figure}[!h]
\begin{center}
\includegraphics[scale=0.38]{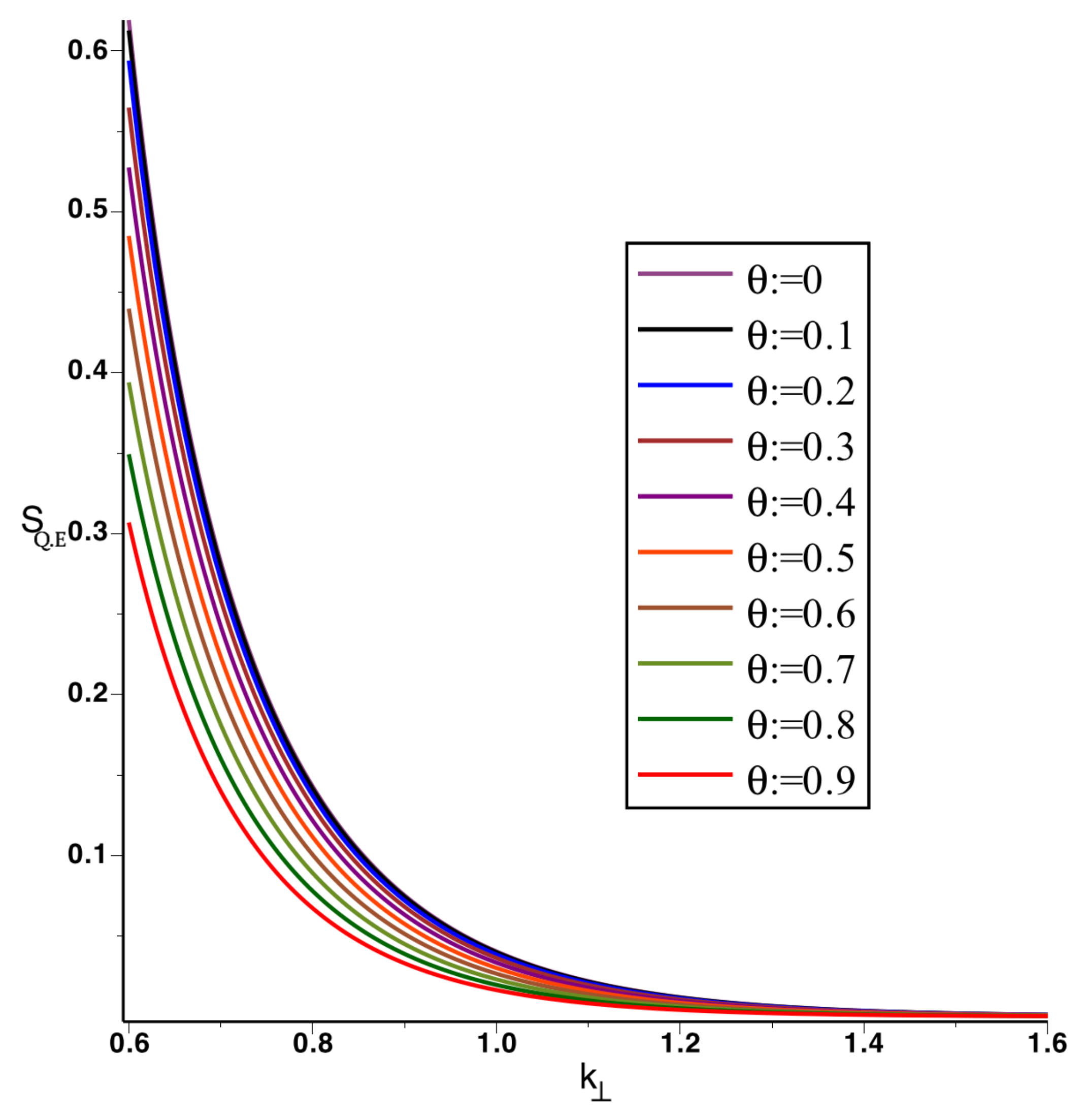} 
\end{center}
\caption{The von Neumann Quantum Entanglement entropy $S_{Q.E}$ as a function of the $k_\perp$-frequency modes  and the noncommutativity $\theta$ parameter}
\label{Fig1}
\end{figure}    
\\Notice that in the fermionic case, as it was pointed out in Ref. \citen{5}, we have two regions, namely the first
is the one corresponding to the out-of-equilibrium state in which the pair creation density has a non thermal behavior. \newpage Furthermore, if we have an anisotropic spacetime, the created particle-antiparticle pair (with the same energy) cannot reach an equilibrium state in all directions, unless their energies exceed a certain critical value $\left(k_{\perp}\approx\frac{1}{2}\right)$ 
beyond which the anisotropic effects become negligible. For $k_{\perp}<\frac{1}{2}$, the particle-antiparticle pair creation velocity (energy) 
is slower than the expansion velocity in the $x$ and $y$ directions and the density of the pair creation is in a non thermal out-of-equilibrium state. 
However, in the bosonic case, we have just one region corresponding to an equilibrium state, starting from $k_{\perp}^{min}$ which is 
similar to the second region in the fermionic case\cite{5}. Moreover, the behavior of $S_{Q.E}$ as a function of $k_{\perp}$ is 
expected, because if $k_{\perp}$ increases, the velocity of the particle's creation increases too, and therefore the information will be 
spread out (decrease of $S_{Q.E}$). Figure 1 illustrates the behavior of $S_{Q.E}$ as a function of $k_{\perp}-$ frequency modes 
for various values of the N.C. $\theta$ parameter. Notice that and contrary to the fermionic case, $S_{Q.E}$ is a decreasing function 
of $\theta$ for a fixed value of $k_{\perp}$. As an example : for $k_{\perp}=0.9$ and when $\theta=0.1$, we see that 
$S_{Q.E}=0.074$, and for $\theta=0.6$ for the same value of $k_{\perp}$, $S_{Q.E}=0.051$ (see FIG. \ref{Fig1}). 
\begin{figure}[!h]
\begin{center}
\includegraphics[scale=0.36]{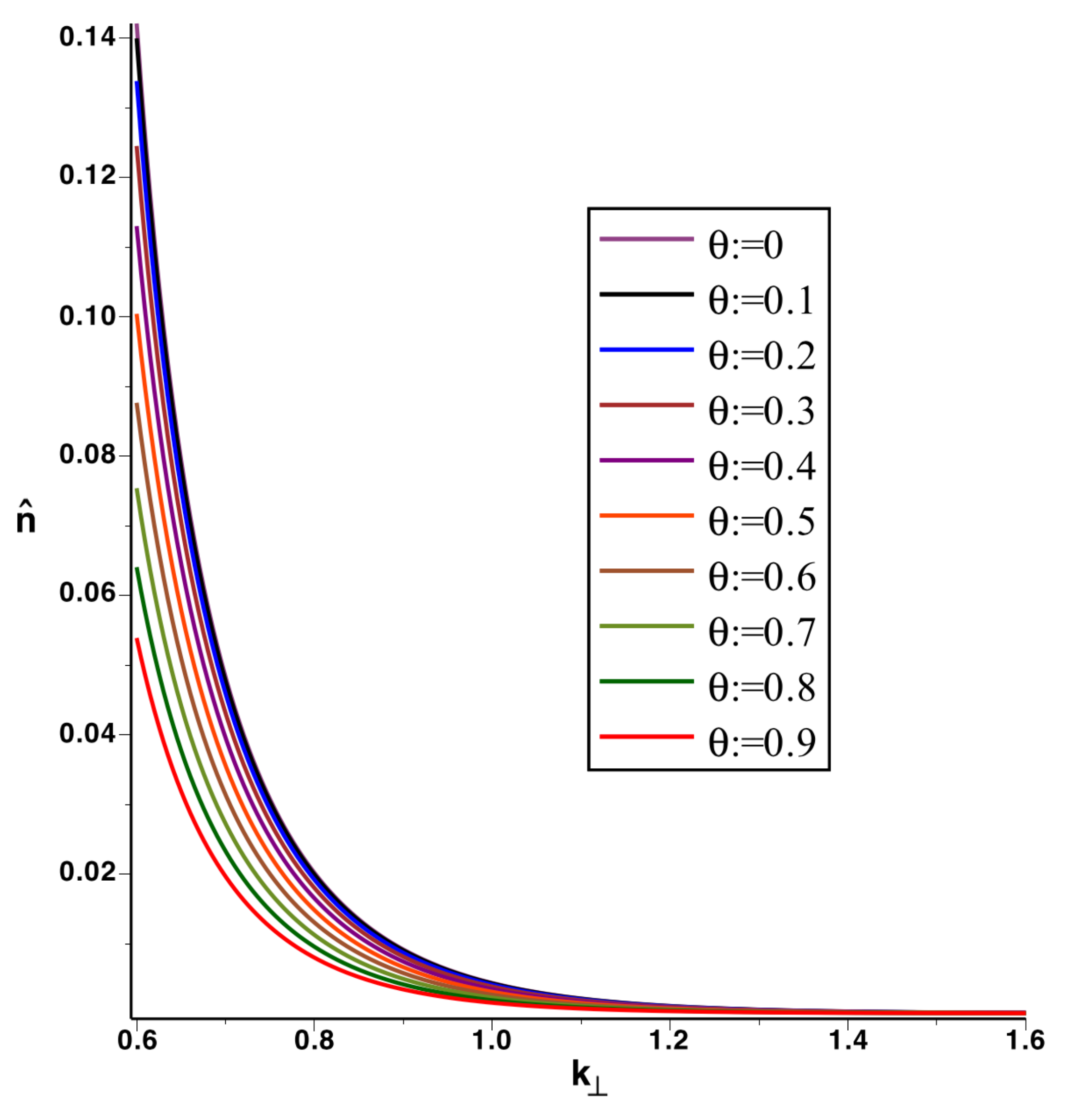} 
\end{center}
\caption{The number density $\hat{n}$ as a function of the $k_\perp$-frequency modes  and the noncommutativity $\theta$ parameter}
\label{Fig2}
\end{figure}
\\In Ref. \citen{5} in region II, the N.C. $\theta$ parameter was interpreted as playing the role of antigravity (e.g. quintessence, dark energy, etc.). 
However, in this paper (bosonic case) as $\theta$ increases, $S_{Q.E}$ decreases. This does not mean that $\theta$ plays the role of gravity. 
In fact, in the bosonic case, the B.E.C. phenomenon which of course depends on $\theta$ increases more than the gravity generated by $\theta$, 
so that overall, the behavior of $S_{Q.E}$ looks like we have gravity. \\Thus, at the end with B.E.C together with antigravitational effects of N.C parameter, $S_{Q.E}$ becomes a decreasing function of $\theta$. Therefore, we can say that $\theta$ generates not only 
antigravity but the B.E.C phenomenon as well. To have a better understanding, let us consider the number density $\hat{n}$ 
(see FIG. \ref{Fig2}). Notice that $\hat{n}$ and $S_{Q.E}$ have the same behavior with respect to $k_{\perp}$ or $\theta$. 
\begin{figure}[!h]
\begin{center}
\includegraphics[scale=0.35]{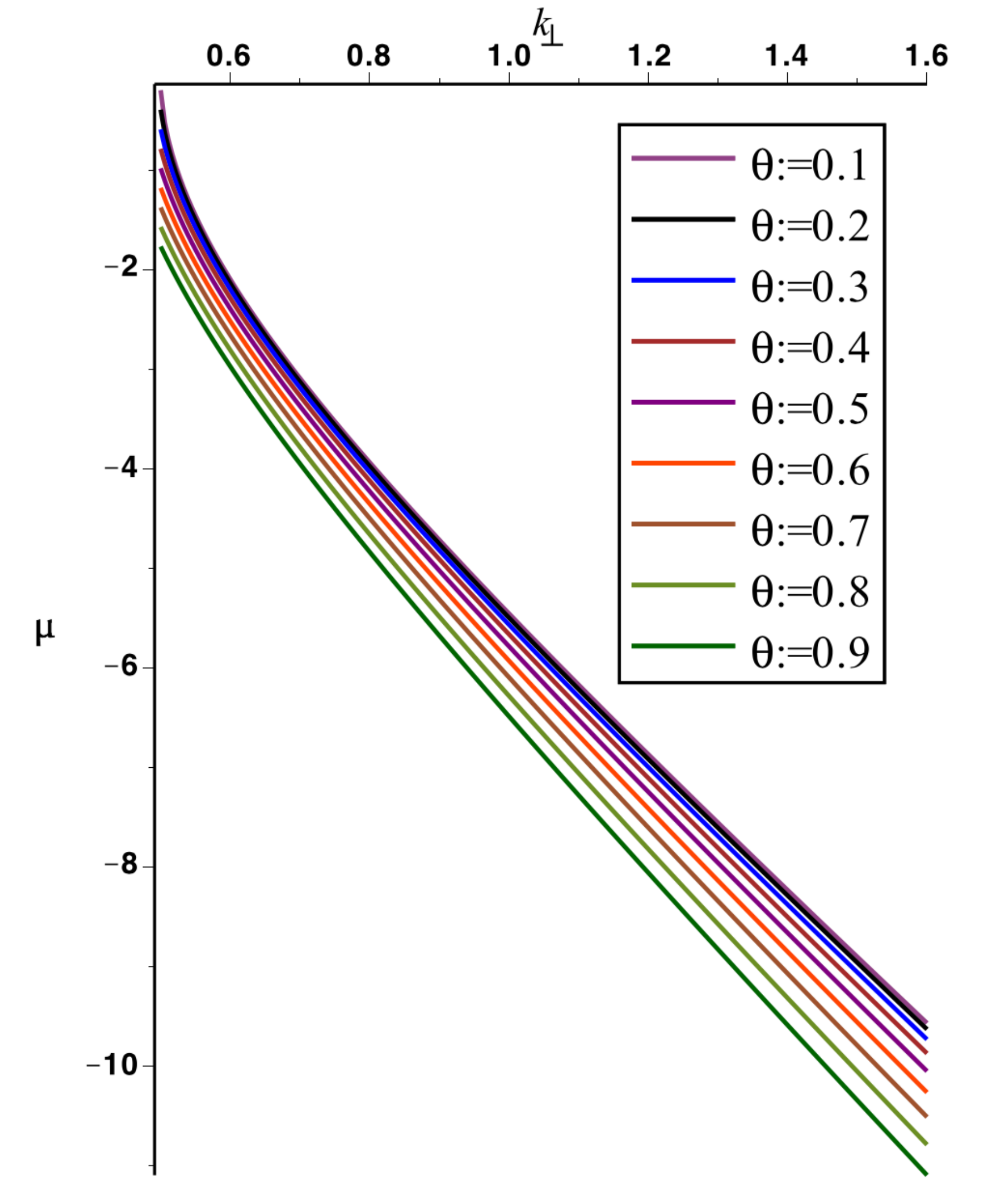} 
\end{center}
\caption{The chemical potential $\mu$ as a function of the $k_\perp$-frequency modes  and the noncommutativity $\theta$ parameter}
\label{Fig3}
\end{figure}
\\ Straightforward calculations show that the chemical potential $\mu$ is proportional to $\log\left(\frac{\hat{n}e^{2\pi\alpha}}
{1+\hat{n}}\right)$,  where $\alpha=\left(1/8\right)\sqrt{25k_{\perp}^{2}\theta^{2}+64k_{\perp}^{2}-16}$ and 
it can be shown to be negative (see FIG. \ref{Fig3}). The reason for $\mu$ being negative is that in the thermodynamical equilibrium where $\mu$ is proportional to 
$-\frac{\Delta S_{Q.E}}{\Delta\hat{n}}$ and $\Delta S_{Q.E}$ is an increasing function of $\Delta\hat{n}$, 
the ratio $\frac{\Delta S_{Q.E}}{\Delta\hat{n}}$ is positive and therefore $\mu$ is negative. Now, if ${\mu}$ 
goes to $0$ implies that $\alpha$ goes to $0$, leading to $k_{\perp}=\frac{4}{\sqrt{25\theta^{2}+64}}$, which means that 
$S_{Q.E}$ goes to $S_{Q.E}^{max}$ (maximally bipartite boson-antiboson entangled state), and the fugacity $z$ goes to $1$. 
In other words, we have a critical point and one has a sort of B.E.C phenomenon. 
\begin{figure}[!h]
        \begin{minipage}{0.5\linewidth}
          %\rule{\linewidth}{0.6\linewidth}  % Only a dummy
          \includegraphics[scale=0.35,bb=120 10 350 550]{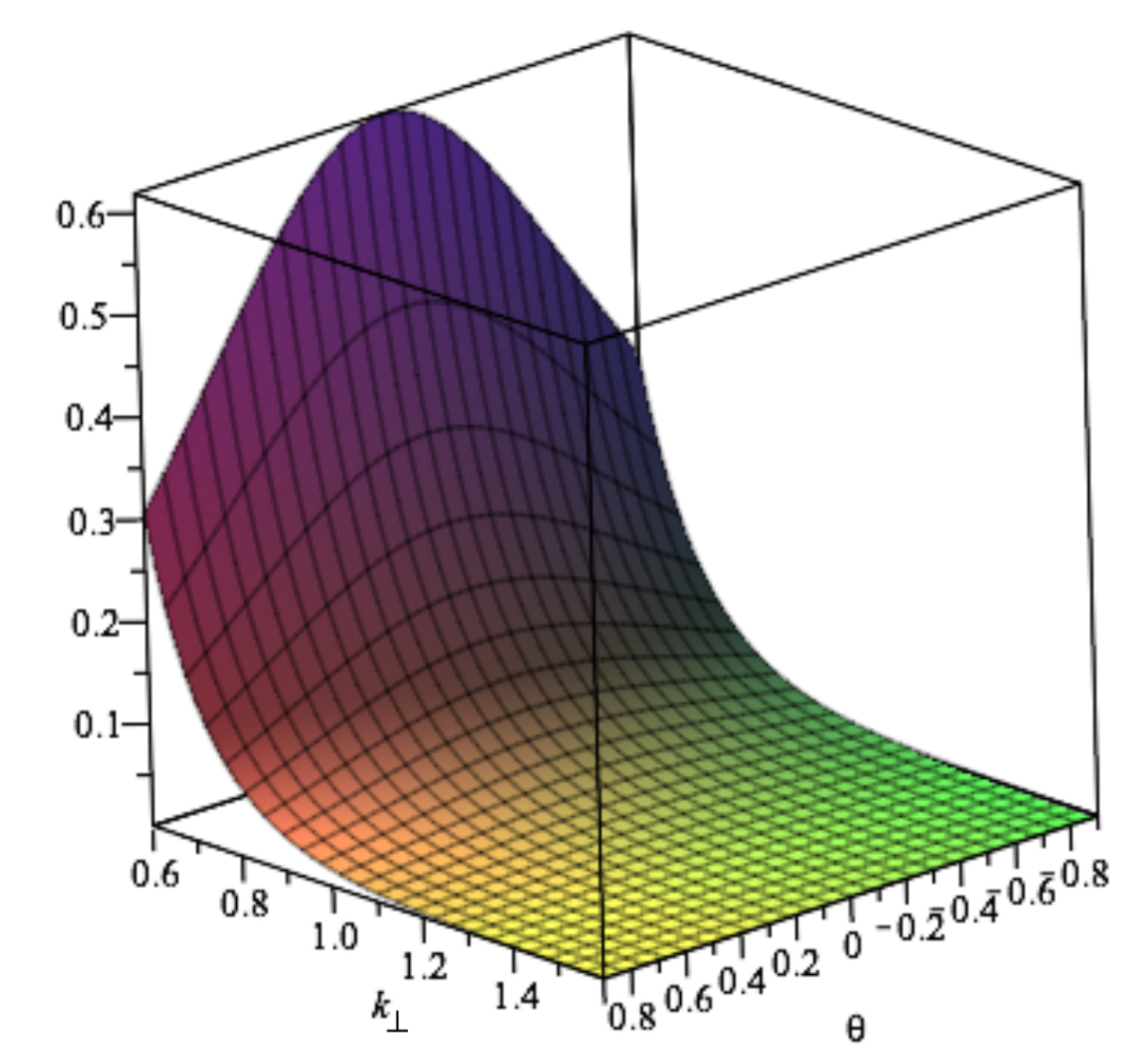} 
          \caption{ $S_{Q.E}$ as a function of the $k_\perp$-frequency modes for various 
values of the N.C.$\theta$ parameter}\label{Fig4}
        \end{minipage}
       %\hfill
        \begin{minipage}{0.4\linewidth}
         %\rule{\linewidth}{0.6\linewidth}  % Only a dummy
          \includegraphics[scale=0.28,bb=80 10 350 550]{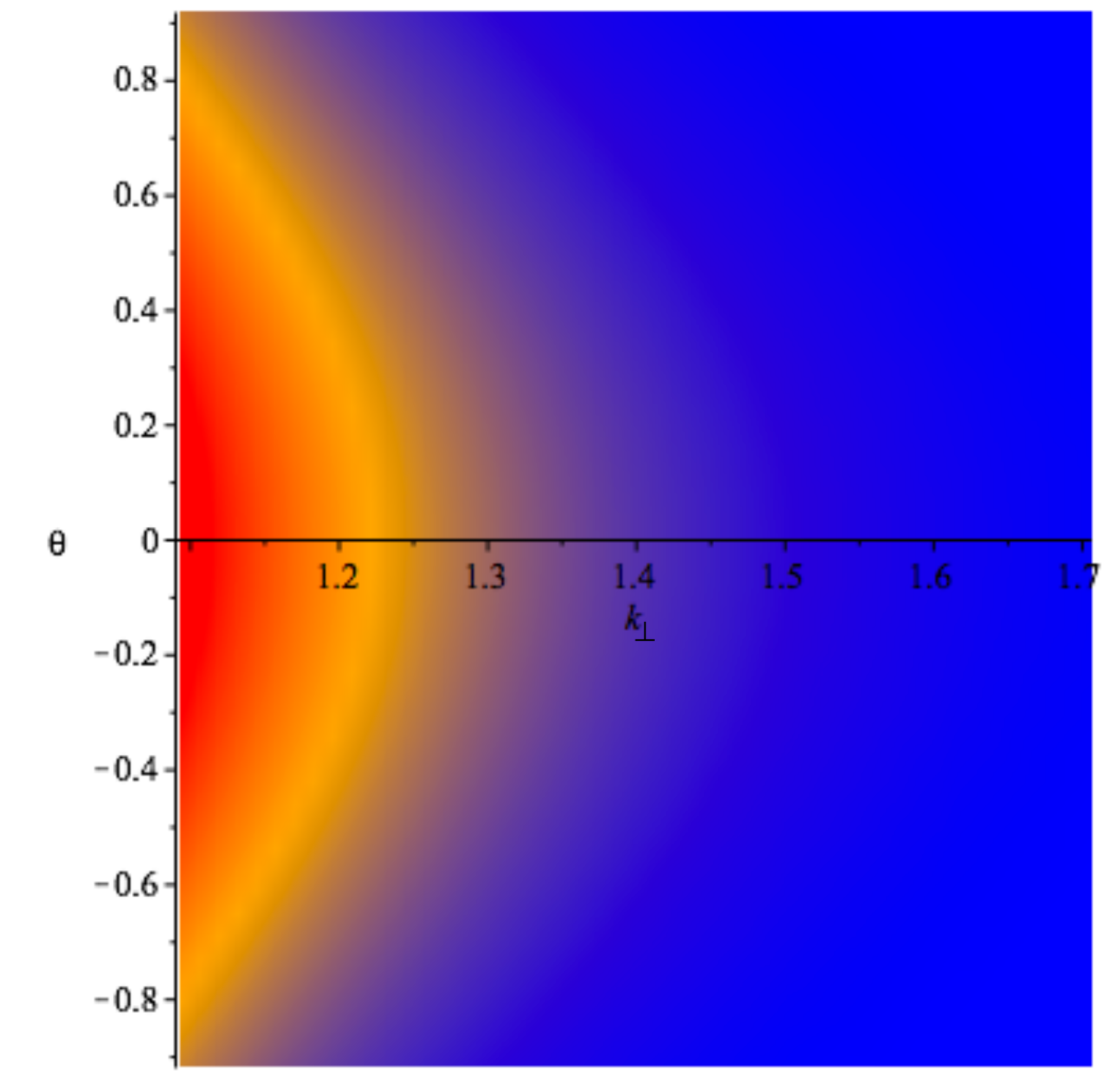}
          \caption{ Density plot of $S_{Q.E}$ as a function of the $k_\perp$-frequency modes for various 
values of the N.C.$\theta$ parameter}\label{Fig5}
        \end{minipage}
      \end{figure}
      %%%%%%%%%%%%%%%%%%%%%%%%%%%%%%%%%%%%%%%%%%%%%%%%%%%%%%%%%%%%%%
      \begin{figure}[!t]
        \begin{minipage}{0.5\linewidth}
          %\rule{\linewidth}{0.6\linewidth}  % Only a dummy
          \includegraphics[scale=0.35,bb=30 10 350 500]{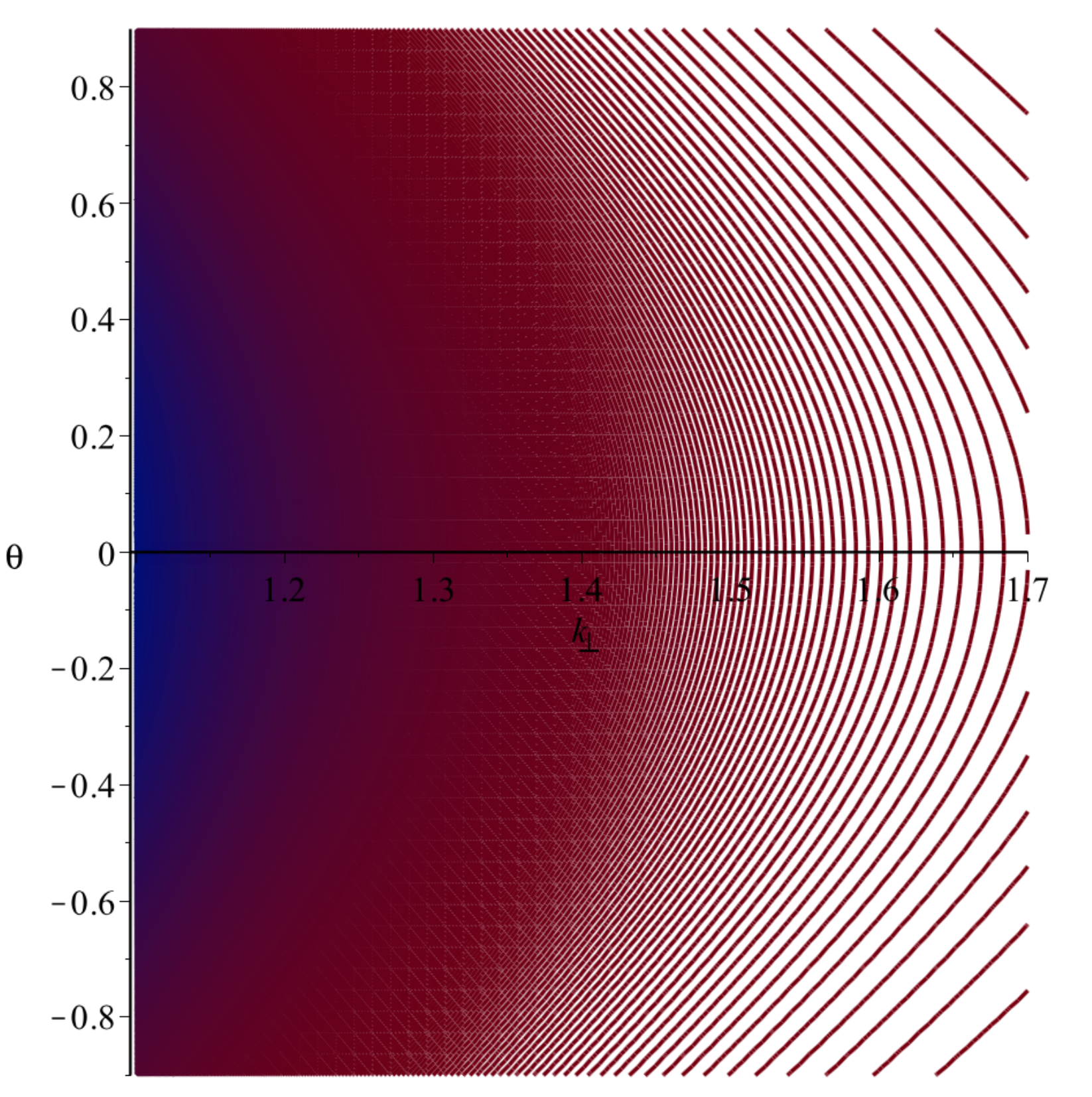} 
          \caption{Contours plot of $\hat{n}$ as a function of the $k_\perp$-frequency modes for various 
values of the N.C.$\theta$ parameter}\label{Fig6}
        \end{minipage}
       %\hfill
        \begin{minipage}{0.4\linewidth}
         %\rule{\linewidth}{0.6\linewidth}  % Only a dummy
          \includegraphics[scale=0.35]{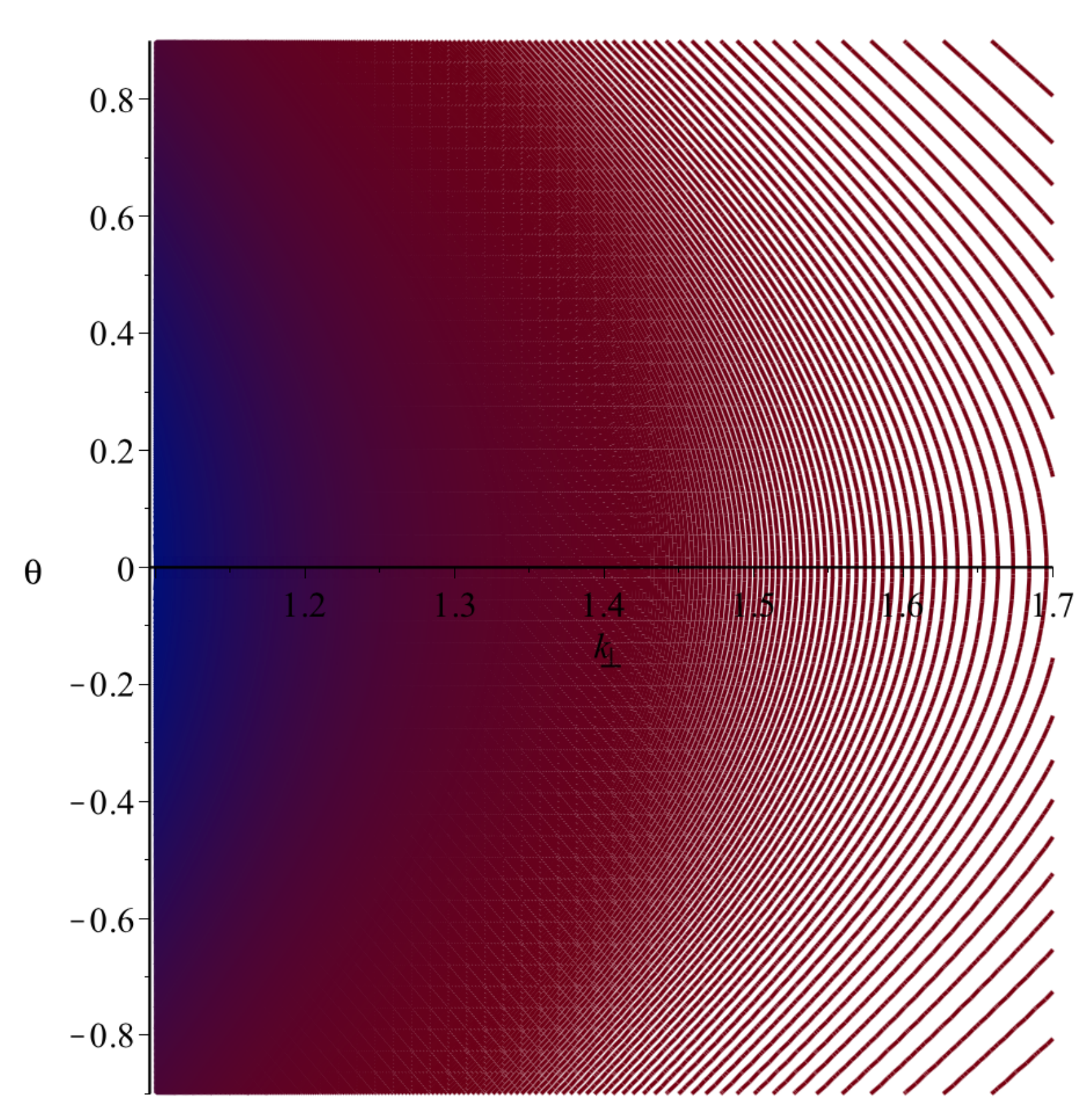}
          \caption{ Contours plot of $S_{Q.E}$ as a function of the $k_\perp$-frequency modes for various 
values of the N.C.$\theta$ parameter}\label{Fig7}
        \end{minipage}
      \end{figure}
\\Figure \ref{Fig4} displays $S_{Q.E}$ as a function of 
$\theta$ and $k_{\perp}$. Figure \ref{Fig5} shows the density plot of $S_{Q.E}$ as a function of $k_{\perp}$ 
for various values of $\theta$.
Figures \ref{Fig6} and \ref{Fig7} illustrate contours plot of the number density $\hat{n}$ and $S_{Q.E}$ as a function of $k_{\perp}$ 
and $\theta$ respectively. 
 Figure \ref{Fig8} represents the chemical potential $\mu$ as a function of $S_{Q.E}^{max}$. \newpage Notice that as $S_{Q.E}^{max}$ 
increases, $\mu$ increases too. The reason is that, if $\mu$ increases the boson-antiboson pair creation rate increases and 
therefore the B.E.C. phenomenon becomes more important leading to an increase in $S_{Q.E}$ (or $S_{Q.E}^{max}$), 
or a decrease of $\theta$ ($S_{Q.E}^{max}$ is a decreasing function of $\theta$).    
\begin{figure}[!h]
\begin{center}
\includegraphics[scale=0.36]{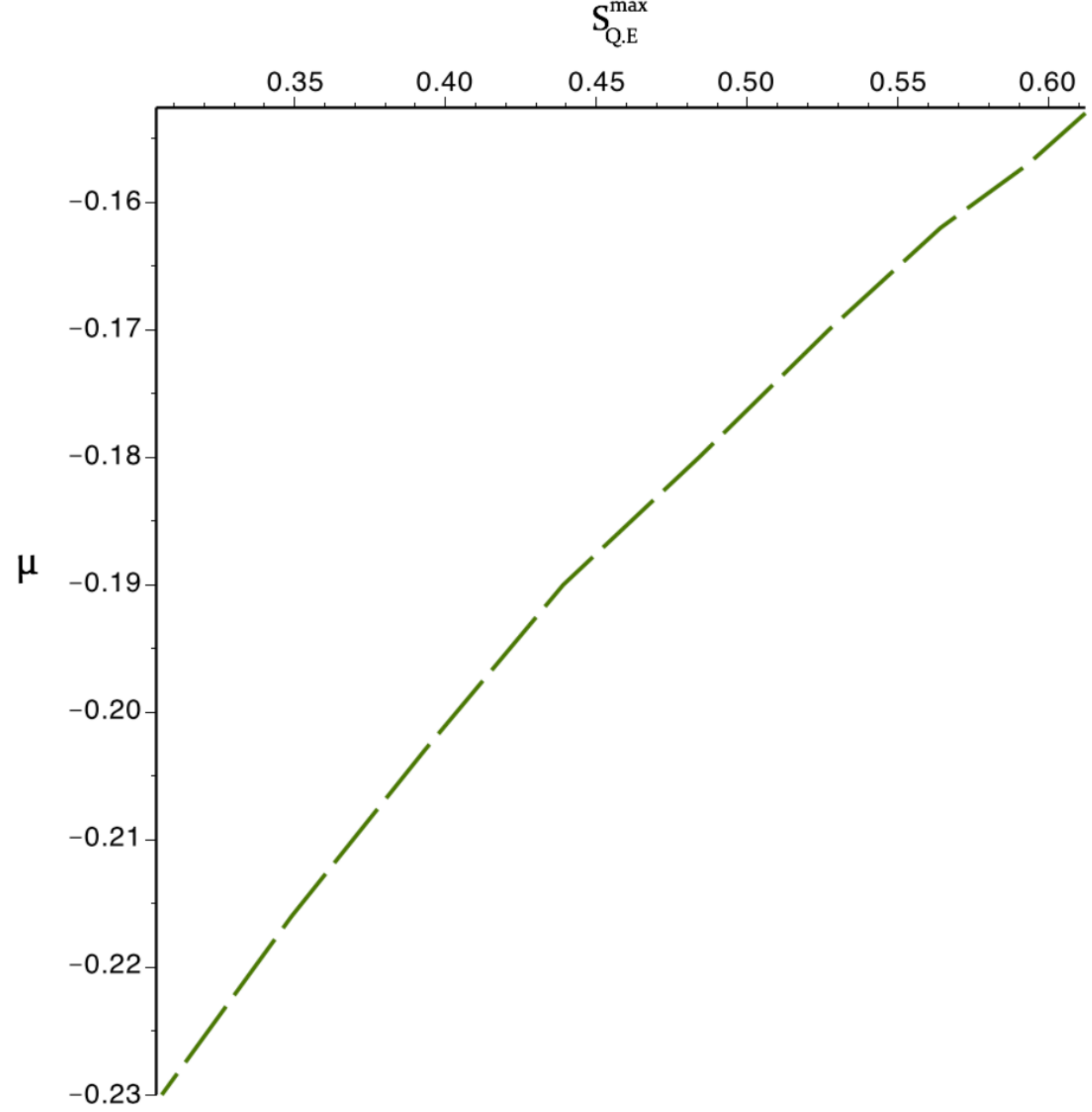} 
\end{center}
\caption{The chemical potential $\mu$ as a function of $S_{Q.E}^{max}$}
\label{Fig8}
\end{figure}\\
   
%%%%%%%%%%%%%%%%%%%%%%%%%%%%%%%%%%%%%%%%%%%%%%%%%%%%%%%%%%%%%%%%%%%%%%%%%%%%%%%%%%%%%%%%%%%%%%%%%%%%%%%%%%%%%%%%%%%%%%%%%%%%%%%%
\newpage
\section{Conclusion}
\label{Conc}
Throughout this paper, we have studied the creation of quantum entanglement between pairs of massless boson-antiboson particles within the framework of the  N.C. Bianchi I 
universe as well as its relationship to thermodynamics. We have first derived the modified N.C. Klein Gordon equation for massless bosons and its solutions. 
Also, due to the complexity of the N.C. anisotropic Bianchi I spacetime structure, the behaviors of $S_{Q.E}$ as a function of $k_{\perp}-$ frequency modes are not 
trivial and different from those obtained in Ref. \citen{16} for the case of isotropic F.R.W. universe (taking a particular solvable case). According to our obtained results, 
the structure and the deformation of the spacetime, as well as the involved particles (fermions or bosons), affect not only the behavior of $S_{Q.E}$ as a function of $k_{\perp}$ but also
the position of $S_{Q.E}^{max}$ as well. It should be noted that, in Ref. \citen{16}, due to the spacetime isotropy, the authors have noticed that for massless bosons one 
has a maximum value of $S_{Q.E}$ independently of the value of $k_{z}$ (not $k_{\perp}$). Our results show that even with massless particles, one can have a non vanishing quantum entanglement 
which depends only on $k_{\perp}= \sqrt{k_{x}^{2}+k_{y}^{2}}$ (because of the Bianchi I spacetime anisotropy and the choice of the N.C. $\theta$ parameter). Similar conclusions were
 obtained in our previous work (Ref. \citen{5}) concerning the fermionic case and compared to Ref. \citen{12}. Thus, the arguments in Refs.\citen{12,16} do not hold in general. 
We have also discussed the behavior of some thermodynamic quantities (like the chemical potential $\mu$) as a function of $k_{\perp}-$ frequency modes and the N.C. $\theta$ parameter. 
We have also shown that the behavior of $S_{Q.E}$ depends on the kind of the involved particles (boson or fermions) during the pair creation process as well as
on the structure and deformation of spacetime. 
Contrary to Ref. \citen{16} and because of the spacetime deformation, the $S_{Q.E}$ for massless bosons does 
not always have a maximum but do depend on the $k_{\perp}$. Notice also that our 
results depend only on the transverse component $k_\perp$ for some allowed values starting from  
$k_{\perp}^{min}=\frac{4}{\sqrt{25\theta^{2}+64}}$ (not the whole components of $k$). 
The maximally entangled state where $S_{Q.E}\rightarrow S_{Q.E}^{max}$ depends strongly on the 
N.C. $\theta$ parameter and knowing the position of $S_{Q.E}^{max}$ one can get information about 
certain thermodynamic quantities (like the chemical potential $\mu$) of the N.C. Bianchi I universe and {\it vice versa}. 
More interesting, we have shown, our numerical results confirm it, that for the minimal value of $k_{\perp}$ we have a 
sort of phase transition (critical point) between the out of equilibrium and equilibrium regions (regions I and II). Furthermore,
 the N.C. $\theta$ parameter plays the role of antigravity and contributes to increase the B.E.C. phenomenon. Finally and unlike our previous result in Ref. \citen{5} 
(where $\vert\theta\vert\leq\frac{8}{5}$), here we cannot get an upper bound of the N.C. $\theta$ parameter.  More studies are under investigation.                                 

%%%%%%%%%%%%%%%%%%%%%%%%%%%%%%%%%%%%%%%%%%%%%%%%%%%%%%%%%%%%%%%%%%%%%%%%%%%%%%%%%%%%%%%%%%%%%%%%%%%%%%%%%%%%%%%%%%%%%%%%%%%%%%%%
\section*{Acknowledgments}
We are very grateful to the Algerian Ministry of Higher Education and  Scientific Research and D.G.R.S.D.T. for financial support.
 This work is also supported by the P.R.F.U. project.
 %%%%%%%%%%%%%%%%%%%%%%%%%%%%%%%%%%%%%%%%%%%%%%%%%%%%%%%%%%%%%%%%
%%%%%%%%%%%%%%%%%%%%%%%%%%%%%%%%%%%%%%%%%%%%%%%%%%%%%%%%%%%%%%%%
\appendix 
\section{N.C. Mathematical Formalism}\label{appendix A}
The N.C. metric is given by (see Refs. \citen{29}, \citen{37}) :
\begin{equation}
\hat{g}_{\mu\nu}=\frac{1}{2}\left(\hat{e}^{b}_{\mu}\ast\hat{e}_{\nu b}+\hat{e}^{b}_{\nu}\ast\hat{e}_{\mu b}\right)
\end{equation}
where, the N.C. Vierbeins $\hat e^{a}_{\mu}$ up to $O(\theta^{2})$ are (see Ref. \citen{5}):
\begin{equation}
\hat e^{a}_{\mu}=e^{a}_{\mu}-i\,\theta^{\nu\rho}\,e^{a}_{\mu\nu\rho}+\theta^{\nu\rho}\,\theta^{\lambda\tau}
\,e^{a}_{\mu\nu\rho\lambda\tau}+O\left( \theta^{3}\right) 
\end{equation}
and
\begin{equation}
e^{a}_{\mu\nu\rho}=\frac{1}{4}\left[ \omega^{ac}_{\nu}\,\partial_{\rho}\,e^{d}_{\mu}+\left( \partial_{\rho}\,\omega^{ac}_{\mu}
+R^{ac}_{\rho\mu}\right) 
e^{d}_{\nu}\right] 
\eta_{cd}
\end{equation}
\begin{eqnarray}
e^{a}_{\mu\nu\rho\lambda\tau}&=&\frac{1}{32}[2\lbrace R_{\tau\nu},R_{\mu\rho}\rbrace^{ab}\,e^{c}_{\lambda}-
\omega^{ab}_{\lambda}(D_{\rho}\,
R^{cd}_{\tau\mu}+\partial_{\rho}\,R^{cd}_{\tau\mu})\,e^{m}_{\nu}\,\eta_{dm} \notag\\ 
&&-\lbrace\omega_{\nu},(D_{\rho}\,R_{\tau\mu}+\partial_{\rho}\,R_{\tau\mu})\rbrace^{ab}\,e^{c}_{\lambda}-\partial_{\tau}
\lbrace\omega_{\nu},(\partial_{\rho}\,\omega_{\mu}+R_{\rho\mu})\rbrace^{ab}\,e^{c}_{\lambda}
\notag\\
&&
-\omega^{ab}_{\lambda}\,\partial_{\tau}(\omega^{cd}_{\nu}\,\partial_{\rho}\,e^{m}_{\mu}+(\partial_{\rho}\,\omega^{cd}_{\mu}
+R^{cd}_{\rho\mu})\,e^{m}_{\nu})\,\eta_{dm}+2\partial_{\nu}\,\omega^{ab}_{\lambda}\,\partial_{\rho}\,\partial_{\tau}\,e^{c}_{\mu}
\notag\\ 
&&
-2\partial_{\rho}
(\partial_{\tau}\,\omega^{ab}_{\mu}+R^{ab}_{\tau\mu})\,\partial_{\nu}\,e^{c}_{\lambda}-\lbrace\omega_{\nu},(\partial_{\rho}
\omega_{\lambda}+R_{\rho\lambda})\rbrace^{ab}\,\partial_{\tau}\,e^{c}_{\mu} \notag\\
&&
-(\partial_{\tau}\,\omega^{ab}_{\mu}+R^{ab}_{\tau\mu})(\omega^{cd}_{\nu}\,\partial_{\rho}\,e^{m}_{\lambda}+(\partial_{\rho}\,
\omega^{cd}_{\lambda}+R^{cd}_{\rho\lambda})\,e^{m}_{\nu}\,\eta_{dm})]\,\eta_{bc} 
\end{eqnarray}
here $R^{ab}_{\mu\nu}$ is the strength field associated with the commutative spin connections $\omega^{ab}_{\mu}$ 
and is defined as:
\begin{equation}
R^{ab}_{\mu\nu}=\partial_{\mu}\,\omega^{ab}_{\nu}-\partial_{\nu}\,\omega^{ab}_{\mu}
+\left( \omega^{ac}_{\mu}\,\omega^{db}_{\nu}-\omega^{ac}_{\nu}\,\omega^{db}_{\mu}\right)\,\eta_{cd}
\end{equation} 
($\eta_{ab}$ is the Minkowski metric).\\
The N.C. spin connections $\hat\omega^{AB}_{\mu}$ up to $O(\theta^{2})$ are:
\begin{equation}
\hat\omega^{AB}_{\mu}=\omega^{AB}_{\mu}-i\,\theta^{\nu\rho}\,\omega^{AB}_{\mu\nu\rho}
+\theta^{\nu\rho}\,\theta^{\lambda\tau}\,\omega^{AB}_{\mu\nu\rho\lambda\tau}+....
\end{equation}
where
\begin{equation}
\omega^{AB}_{\mu\nu\rho}=\frac{1}{4}\lbrace\omega_{\nu},\partial_{\rho}\,\omega_{\nu}+R_{\rho\mu}\rbrace^{AB}
\end{equation}
%%%%%%%%%%%%%%%%%%%%%%%%%%%%%%%%%%%%%%%%%
\begin{eqnarray}
\omega^{AB}_{\mu\nu\rho\lambda\tau}&=&\frac{1}{32}(-\lbrace\omega_{\lambda},\partial_{\tau}\lbrace\omega_{\nu},
\partial_{\rho}\,\omega_{\mu}
+R_{\rho\mu}\rbrace\rbrace+2\lbrace\omega_{\lambda},\lbrace R_{\tau\nu},R_{\mu\rho}\rbrace\rbrace\notag\\
&&
-\lbrace\omega_{\lambda},\lbrace\omega_{\nu},D_{\rho}\,R_{\tau\mu}+\partial_{\rho}\,R_{\tau\mu}\rbrace\rbrace
-\lbrace\lbrace\omega_{\nu},\partial_{\rho}\,\omega_{\lambda}+R_{\rho\lambda}\rbrace,(\partial_{\tau}\,\omega_{\mu}
+R_{\tau\mu})
\rbrace\notag\\
&&
+2[\partial_{\nu}\,\omega_{\lambda},\partial_{\rho}(\partial_{\tau}\,\omega_{\mu}
+R_{\tau\mu})])^{AB}
\end{eqnarray}
%%%%%%%%%%%%%%%%%%%%%%%%%%%%%%%%%%%
here
\begin{equation}
\lbrace\alpha,\beta\rbrace^{AB}=\alpha^{AC}\,\beta^{B}_{C}+\beta^{AC}\,\alpha^{B}_{C}
\end{equation}
\begin{equation}
[\alpha,\beta]^{AB}=\alpha^{AC}\,\beta^{B}_{C}-\beta^{AC}\,\alpha^{B}_{C}
\end{equation}
and
\begin{equation}
D_{\mu}\,R^{AB}_{\rho\sigma}=\partial_{\mu}\,R^{AB}_{\rho\sigma}+(\omega^{AC}_{\mu}+
R^{DB}_{\rho\sigma}+\omega^{BC}_{\mu}\,R^{DA}_{\rho\sigma})\,\eta_{CD}
\end{equation}
%%%%%%%%%%%%%%%%%%%%%%%%%%%%%%%%%%%%%%%%%%%%%%%%%%%%%%%%%%%%%%%%
We propose the following action:
\begin{equation}
S=\frac{1}{2{k}^{2}}\int d^{4}x\left( \mathcal{L}_{G}+ \mathcal{L}_{SC}\right)
\end{equation} 
where $\mathcal{L}_{G}$ and $\mathcal{L}_{SC}$ stand for the pure gravity and matter scalar densities corresponding to the charged scalar particle.\\
We define 
\begin{equation}
\mathcal{L}_{G}=\hat{e}\ast\hat{R}
\end{equation}
and 
\begin{equation}
\mathcal{L}_{SC}=\hat{e}\ast\left(\hat{g}^{\mu\nu}\ast\left(\hat{D}_{\mu}\hat{\varphi}\right)^{\dagger}\ast\hat{D}_{\nu}\hat{\varphi}\right)
\end{equation}
with :
\begin{equation}
\hat{R}=\hat{e}^{\mu}_{\ast a}\ast\hat{e}^{\nu}_{\ast b}\ast\hat{R}^{ab}_{\mu\nu}
\end{equation}
Applying the principle of least action, it is easy to show that the modified field equations are given by:
\begin{equation}
\frac{\partial\mathcal{L}}{\partial\hat{\varphi}}-\partial_{\mu}\frac{\partial\mathcal{L}}{\partial\left(\partial_{\mu}\hat{\varphi}\right)}+\partial_{\mu}\partial_{\nu}\frac{\partial\mathcal{L}}{\partial\left(\partial_{\mu}\partial_{\nu}\hat{\varphi}\right)}-\partial_{\mu}\partial_{\nu}\partial_{\sigma}\frac{\partial\mathcal{L}}{\partial\left(\partial_{\mu}\partial_{\nu}\partial_{\sigma}\hat{\varphi}\right)}+O\left( \theta^{3}\right)=0
\end{equation} 
%%%%%%%%%%%%%%%%%%%%%%%%%%%%%%%%%%%%%%%%%%%%%%%%%%%%%%%%%%%%%%%%%%%%%%%%%%%%%%%%%%%%%%%%%%%%%%%%%%%%%%%%%%%%%%%%%%%%%%%%%%%%%%%%

%%%%%%%%%%%%%%%%%%%%%%%%%%%%%%%%%%%%%%%%%%%%%%%%%%%%%%%%%%%%%%%%
\end{document}